\documentclass[12pt]{article}
\usepackage[dvips]{graphicx}
\usepackage{amssymb,amsmath}

\begin{document}
\begin{center}
{  \Large  {\bf Squeezed coherent states for gravitational well in noncommutative space
\\\
 } }
\end{center}
\vskip 0pt
\begin{center}
{\it {\large $P \hskip 2 pt Patra$\footnote { e-mail:
monk.ju@gmail.com}   }\\
\it{Department of Physics, Brahmananda Keshab Chandra College,\\ Kolkata, India-700108}\\ $ $ \\

\it {\large $J \hskip 2 pt P \hskip 2 pt Saha$\footnote {e-mail:jyotiprasadsaha@gmail.com} } \\{Department of Physics ,
                                                                             University of Kalyani, \\
                              Kalyani, Nadia, West Bengal,  
                                    India - 741235} 
                                    \\$ $\\
                                    \it{ \large $ K \hskip 2 pt Biswas $
                                    \footnote{Corresponding author) e-mail:klpnbsws@gmail.com}}\\  
                                   {Department of Physics ,
                                                                                                                University of Kalyani, \\
                                                                                                                                              Kalyani, Nadia, West Bengal,  
                                                                                                                                                    India - 741235} \\
                                  {Department of Physics, Sree Chaitanya College, Habra, \\North 24 Parganas, West Bengal, India-743268} }
                                  \end{center}

                                  \vskip 20pt
                                  
 \begin{center}
 {\bf Abstract}
 \end{center}
 \par Gravitational well is a widely used system for the verification of the quantum weak equivalence principle (WEP).  We have studied the quantum gravitational well (GW) under the shed of non-commutative (NC) space so that the results can be utilized for testing the validity of WEP in NC-space. To keep our study widely usable, we have considered both position-position and momentum-momentum non-commutativity.\\
 Since coherent state (CS) structure provides a natural bridge between the classical and quantum domain descriptions, the quantum domain validity of purely classical phenomena like free-fall under gravity might be verified with the help of CS.  We have constructed CS with the aid of a Lewis-Riesenfeld phase space invariant operator.  We deduced the uncertainty relations from the expectation values of the observables and shown that the solutions of the time-dependent Schr\"{o}dinger equation are squeezed-coherent states.\\
  \par keywords: Coherent state; Lewis-Riesenfeld invariance; Non-commutative geometry; Gravitational well
 \par PACS: 03.65.Sq, 03.65.Fd, 04.60.−m, 02.90.+p, 04.20.Jb
 
 \section{Introduction }
 \noindent It is almost a Gospel that the fundamental concept of space-time is mostly compatible with quantum theory in non-commutative space \cite{Witten,Nikita,Richard,Hull,Bertolami,ncexpt1,energydependentparameter,nc1,nc2,nc3,nc4,nc5,nc6,nc7,nc8,nc9,nc10}. However, despite the numerous proposals, the unified quantum theory of gravity lacks any direct experimental evidence. In particular, most of the theories of quantum gravity appears to predict departures from classical relativity only at energy scales on the order of $10^{19}$ GeV (By way of comparison, the LHC was designed to run at a maximum collision energy of $14$  TeV \cite{lhc}).  It is not difficult to foretell that the attainable energy in colliders will not satisfy the required energy level shortly.
 Instead of building a larger collider, the energy limitation problem can be overcome with the help of performing passive experiments using interferometry. Numerous proposals are there in this regime 
 \cite{Hamil:2020zvy,Merad:2019sxl,Ching:2019frl,Santos:2019uat,Meljanac:2019qzw,Jentschura:2018mlv,Dey:2018clp,Khodadi:2018scn,Khodadi:2018kqp,Dey:2018bmr}. However, to the best of knowledge of present authors, none of them had yet been carried out.\\
 Since optical coherence can be seen in Young's double-slit experiment.  It maximizes the contrast in the interference pattern. Therefore, the use of Fock and coherent states might improve the experimental scope of non-commutative space. \cite{Dey:2018clp,Khodadi:2018scn,Khodadi:2018kqp,Dey:2018bmr,Chattopadhyay:2017agv,interfero1,interfero2,interfero3}.  In particular,
 one may use Fock and coherent states, which describe the electromagnetic input field, a multi-photon counting apparatus to a multi-slit Young’s experiment to perform such experiments. Since the aims and scope of the present article are limited to the mathematical construction of coherent states, we are not going to the detailed study in the experimental regime.  However, one may consult for experimental scope from this perspective. \cite{Dey:2018clp,Khodadi:2018scn,Khodadi:2018kqp,Dey:2018bmr,Chattopadhyay:2017agv,interfero1,interfero2,interfero3,Nieto,Simmons,Hartley,Ray,Pedrosa}.
 \\
 In this article, we have utilized the Lewis-Riesenfeld phase-space invariant method (LRM)for the construction of the coherent state structure of the system under consideration. LRM is based on the construction of an invariant operator (IO) in phase space corresponding to a time-dependent Hamiltonian $\hat{H}(t)$ \cite{Lewis,Lewis1,Lewis2,Lewis3,Lewis4,Lewis5,Lewis6,Lewis7}.   Up to a time-dependent phase factor, the eigenstates of IO $(\hat{\mathcal{I}}(t))$ will satisfy the time-dependent Schr\"{o}dinger equation for $\hat{H}(t)$. However, $(\hat{\mathcal{I}}(t))$ and $\hat{H}(t)$ will not be isospectral in general. Specifically, the eigenvalues of $(\hat{\mathcal{I}}(t))$ are time-independent, whereas the same may be time-dependent for $\hat{H}(t)$.
 LRM is an efficient tool to deal with the exact solutions for time-dependent Hamiltonians. However, one can utilize LRM for time-independent problems as well \cite{Penna,Bagrov,Torre,Guerrero1,Geloun,Maamache}. One can incorporate a time-dependent parameter in such a manner that the time-dependency goes off at some limit \cite{Penna,Bagrov,Torre,Guerrero1,Geloun}.  One can even use the route of the direct approach without any artificial time-dependent parameter \cite{Maamache,pinaki}. 
 The connection between the LRM and Bathe Ansatz  method (BAM) can be found in \cite{dongbook}. One can obtain the constraints between the undetermined parameters using BAM. \\
 In the present article,  the direct method is utilized to construct Lewis-Riesenfeld invariant operator (LRIO) corresponding to the Hamiltonian of a  gravitational well in non-commutative space. 
 The importance of free-falling particle under gravity in noncommutative space including the aspects of weak equivalence principle may be found in \cite{Zych,Will,Rosi,Aquilera,Schlippert,Haugan,Geiger}.We shall see in what follows that since our constructed IO is linear in momentum, we can construct an annihilation operator with the help of IO $(\hat{\mathcal{I}}(t))$. Construction of eigenstates of $(\hat{\mathcal{I}}(t))$ then enables us to have coherent states for the concerned scenario. \\
 The organization of the article is the following. After a brief description of the system under consideration, we have shown the existence of a Lewis-Riesenfeld phase-space invariant operator for the gravitational well in noncommutative space. Subsequently, the construction of the coherent state structure along with the expectation-values are explicitly mentioned.
 \section{System under consideration: Free falling under gravity in noncommutative space}
 \noindent Since we are dealing with the free-falling under gravity in noncommutative space (NCS), without loss of generality we can confine ourselves in two spatial dimensions in NCS, namely $x'$ and $y'$. Let us choose our $x'$ axis in the direction of the attraction due to gravity. The dynamics for the $y'$ direction in NCS remains something like free-particle. We aim to write down the coherent state structure for this system. \\
 Better or worse, we are adopting the usual technique of writing the quantum version of a theory corresponding to a known classical dynamical system with the aid of the Bohr-correspondence principle. One may wonder why most of the computations of quantum theory can not stand on its ground without the help of its corresponding known classical dynamical system. One may even demand the justification of the applicability of the direct promotion of classical theory to its quantum version for NCS. Keeping aside these foundational issues, let us concentrate on the aims and scope of the present article and write down the energy operator (Ḫamiltonian) in NCS as follows.
 \begin{equation}\label{nc hamiltonian}
 \hat{H}_{nc}=\frac{\hat{p}_x'^2}{2m} + \frac{\hat{p}_y'^2}{2m} + mg\hat{x}' .
 \end{equation}
 Where the mass of the particle and the acceleration due to gravity is denoted by $m$ and $g$ respectively. $\hat{p}_x'$ and $\hat{p}_y'$ are the conjugate momentum operators corresponding to $\hat{x}'$ and $\hat{y}'$ respectively. We are considering both position-position and momentum-momentum noncommutativity to keep our discussion fairly general. 
 Following commutation relations for variables $\left\{ x',y',p_x',p_y'\right\}$ in NCS are utilized in present article.
 \begin{eqnarray}\label{NCcommutation}
 \left[ x',y' \right]=i\theta , \nonumber \\
 \left[ p_x',p_y' \right]=i\eta ,  \\
 \left[ x_i',p_j' \right]=i\hbar_{eff}\delta_{ij}. \nonumber 
 \end{eqnarray}
 Where 
 \begin{eqnarray}
 \hbar_{eff}=(1+\zeta)\hbar, \\
 \mbox{with}\; \zeta =\frac{\theta\eta}{4\hbar^2}. \nonumber
 \end{eqnarray}
 $\delta_{ij}$ are Kronecker delta with the properties
 \begin{eqnarray}
 \delta_{ij}=\left\{\begin{array}{ccc}
 1 &   \mbox{for} & i=j \\
 0 &   \mbox{for} & i\neq j
 \end{array}
 \right. .
 \end{eqnarray}
 It is worth noting that, since  the non-commutative parameters $\theta (E)$ and $\eta (E)$ are energy dependent, the effective Planck constant ($\hbar_{eff}$) might be energy-dependent as well.  However, this setup of energy dependency is consistent with
 the usual commutative space-time quantum mechanics only if $\zeta=\frac{\theta\eta}{4\hbar^2} \ll 1$ \cite{zeta1,zeta2,zeta3}. 
 Indeed, one can recover the structure of classical commutative space for quantum mechanics by setting the parameters $\theta$ and $\eta$ to zero. If one wishes to be confined in only position-position noncommutativity then $\eta$ has to be set to zero.\\
 Since our usual notion of calculus are mentally and practically settled in commutative space, it will be a wise decision if we could transform the whole problem to some equivalent commutative space structure. This can be done by the following transformation of co-ordinates.
 \begin{eqnarray}\label{transformation c to nc}
 \left(\begin{array}{c}
 x' \\
 y'\\
 p_x'\\
 p_y'
 \end{array}\right)= 
 \left(\begin{array}{cccc}
 1 & 0 & 0 & -\frac{\theta}{2\hbar} \\
 0 & 1 & \frac{\theta}{2\hbar} & 0\\
 0 & \frac{\eta}{2\hbar}& 1 & 0\\
 -\frac{\eta}{2\hbar}& 0 & 0& 1
 \end{array}\right)
 \left(\begin{array}{c}
 x \\
 y \\
 p_x \\
 p_y
 \end{array}\right).
 \end{eqnarray}
 $\left\{ x,y,p_x,p_y\right\}$ are usual co-ordinates in classical commutative space in which the commutation relations are given by
 \begin{eqnarray}\label{Ccommutation}
 \left[ x,y \right]=\left[ p_x,p_y \right] =0 , \nonumber \\
 \left[ x_i,p_j \right]=i\hbar\delta_{ij}. 
 \end{eqnarray}
 For computational purpose, with the help of ~\eqref{transformation c to nc} we can now write down the equivalent quantum hamiltonian for the system under consideration (~\eqref{nc hamiltonian}) in terms of classical commutative space operators  as follows.
 \begin{eqnarray}\label{c hamiltonian}
 \hat{H}_c = \frac{\hat{p}_x^2}{2m} + \frac{\hat{p}_y^2}{2m} + \frac{\eta}{2m\hbar}\left(\hat{y}\hat{p}_x - \hat{x}\hat{p}_y\right) + mg \left( \hat{x}-\frac{\theta}{2\hbar}\hat{p}_y\right) + \frac{\eta^2}{8m\hbar^2}\left(\hat{x}^2 +\hat{y}^2\right).
 \end{eqnarray}
 In the next section we have constructed an invariant operator(IO) which is utilized to construct the coherent state structure of the system.
 \section{Construction of Lewis-Riesenfeld invariant operator and coherent state structure}
 An operator  $ \hat{\mathcal{I}}(t)$ is invariant means
 \begin{equation}\label{invariant condition}
 \dot{\hat{\mathcal{I}}}(t) =\frac{\partial \hat{\mathcal{I}}(t)}{\partial t} + \frac{1}{i\hbar} \left[\hat{\mathcal{I}}(t), \hat{H}\right] =0 .
 \end{equation}
 Here dot $(^.)$ denotes total derivative with respect to time. We shall use this shorthand notation throughout this article unless otherwise specified.
 The existence of a close form LR-invariant $\hat{\mathcal{I}}(t)$ based on the existence of a finite number of generators $(\hat{\mathcal{O}}_i)$ of the quasi-algebra with respect to the Hamiltonian $\hat{H}$ such that the equation \eqref{invariant condition} is satisfied. Specifically, we are looking for an invariant operator of the form 
 \begin{equation}
 \hat{\mathcal{I}}(t) =\sum_{j=0}^N \mu_{j} (t) \hat{\mathcal{O}}_j , 
 \end{equation}
 such that the following quasi-algebra is closed for finite $N$.
 \begin{equation}
 \left[\hat{H}, \hat{\mathcal{O}}_i\right]= \sum_{k=1}^{N} \nu_{kj} \hat{\mathcal{O}}_j; \;\; i=1,...N.
 \end{equation}
 Where $\nu_{kj}$ are the structure constants of the algebra and  $\mu_{j}$ are arbitrary functions of time. One can note that a quasi-algebra can be generated by the following set of generators 
 \begin{eqnarray}\label{generators}
 \hat{\mathcal{O}}_1=\hat{x},\;\; \hat{\mathcal{O}}_2=\hat{y}, \nonumber \\
 \hat{\mathcal{O}}_3=\hat{p}_x,\;\; \hat{\mathcal{O}}_4=\hat{p}_y.
 \end{eqnarray}
 One can identify the following closed quasi-algebra.
 \begin{eqnarray}\label{quasialgebra}
 \left[\hat{H}_c,\hat{x}\right]&=&-i\frac{\hbar}{m}\hat{p}_x -i\frac{\eta}{2m}\hat{y}, \\
 \left[\hat{H}_c,\hat{y}\right]&=&-i\frac{\hbar}{m}\hat{p}_y +i\frac{\eta}{2m}\hat{x}+\frac{i}{2}mg\theta \hat{\mathbb{I}}, \\
 \left[\hat{H}_c,\hat{p}_x\right]&=&-i\frac{\eta}{2m}\hat{p}_x +i\frac{\eta^2}{4m\hbar}\hat{x}+img\hbar \hat{\mathbb{I}}, \\
 \left[\hat{H}_c,\hat{p}_y\right]&=&i\frac{\eta}{2m}\hat{p}_x +i\frac{\eta^2}{4m\hbar}\hat{y}.
 \end{eqnarray}  
 Where $\hat{\mathbb{I}}$ is the identity operator.\\
 Now, one can consider an invariant operator ($\hat{\mathcal{I}}(t)$) of the form
 \begin{equation}\label{invariantansatz}
 \hat{\mathcal{I}}(t)=A(t)\hat{p}_x+B(t)\hat{p}_y+C(t)\hat{x}+D(t)\hat{y}+ \alpha(t)\hat{\mathbb{I}}.
 \end{equation}
 Using the ansatz ~\eqref{invariantansatz} in ~\eqref{invariant condition}, we get the following set of coupled equations.
 \begin{eqnarray}\label{coupledABCD}
 \frac{d}{dt}\left(\begin{array}{c}
 A \\
 B\\
 C\\
 D\\
 \alpha
 \end{array}\right)= 
 \left(\begin{array}{ccccc}
 0 & -\frac{\eta}{2m\hbar} & \frac{1}{m} & 0 & 0 \\
 \frac{\eta}{2m\hbar} & 0 & 0 & \frac{1}{m} & 0 \\
 -\frac{\eta^2}{4m\hbar^2} & 0 & 0 & -\frac{\eta}{2m\hbar} & 0 \\
 0 &  -\frac{\eta^2}{4m\hbar^2}& \frac{\eta}{2m\hbar} & 0 & 0\\
 mg & 0 & 0 & \frac{mg\theta}{2\hbar} & 0
 \end{array}\right)
 \left(\begin{array}{c}
 A \\
 B \\
 C \\
 D\\
 \alpha
 \end{array}\right)= 
 \left(\begin{array}{c}
 0 \\
 0\\
 0\\
 0\\
 0
 \end{array}\right).
 \end{eqnarray}
 By solving ~\eqref{coupledABCD}, we can see that   $A(t),\; B(t),\; C(t), \;D(t)$ and $\alpha(t)$ are completely expressible in terms of any two of them. For example, we can express them in terms of $A(t)$ and $B(t)$, which can explicitly be written down  as follows.
 \begin{eqnarray}
 A(t)&=&-\frac{1}{m\omega}A_0+\frac{1}{\omega}\left(B_1\sin \omega t -B_2\cos\omega t\right), \\
 B(t)&=&\frac{1}{m\omega}B_0+ \frac{1}{\omega}\left(B_1\cos \omega t + B_2\sin\omega t\right),
 \end{eqnarray}
 with
 \begin{equation}
 \omega = \frac{\eta}{m\hbar}.
 \end{equation}

 The integration constants $A_0,B_0,B_1,B_2 $ are complex numbers in general.\\
 Upto a time-dependent phase factor $e^{i\nu_\lambda(t)}$, $\mathcal{I}(t)$ and $\hat{H}_c$ share same eigen-functions.  To obtain the eigen-function of $\mathcal{I}(t)$, we can solve the following eigen-value equation with time independent eigen-value $\lambda$.
 \begin{equation}\label{eigenvalueequation of phi}
 \hat{\mathcal{I}}\Phi_\lambda =\lambda \Phi_\lambda.
 \end{equation}
 In position representation the solution of ~\eqref{eigenvalueequation of phi} is given by
 \begin{eqnarray}\label{Phiform}
 \Phi_\lambda (x,y,t)=\Phi_0 \exp\left\{\frac{(\lambda -\alpha)i}{2\hbar}\left(\frac{x}{A}+\frac{y}{B}\right) -\frac{i}{2\hbar}\left(\frac{C}{A}x^2 +\frac{D}{B}y^2\right)\right\}.
 \end{eqnarray}
 Now the eigen-function of $\hat{H}_c$ can be written in the form
 \begin{equation}\label{psilambda}
 \psi_\lambda (x,y,t)=e^{i\nu_\lambda(t)}\Phi_\lambda (x,y,t).
 \end{equation}
 $\nu_\lambda(t)$ then satisfies the following equation
 \begin{equation}\label{nuequation}
 \hbar \dot{\nu}_\lambda(t)\Phi_\lambda (x,y,t) =\left[i\hbar \frac{\partial}{\partial t}-H\right] \Phi_\lambda (x,y,t).
 \end{equation}
 Using ~\eqref{Phiform} and ~\eqref{c hamiltonian} in ~\eqref{nuequation}, we have obtained the following differential equation for $\nu_\lambda(t)$.
 \begin{eqnarray}\label{nuequationexplicit}
 \hbar \dot{\nu}_\lambda = i\hbar \frac{\dot{\Phi_0}}{\Phi_0}- \frac{i\hbar}{2m} \left(\frac{C}{A} + \frac{D}{B} \right) + \frac{mg\theta}{2\hbar}\frac{\left(\lambda - \alpha\right)}{B}  \nonumber \\ - \frac{1}{8m}\left(\lambda - \alpha\right)^2\left(\frac{1}{A^2}+\frac{1}{B^2}\right) .
 \end{eqnarray}
 Since the phase factor $\nu_\lambda(t)$ is a function of $t$ only, the right hand side of ~\eqref{nuequationexplicit} should be a function of $t$ only. This restricts the choices of the coefficients in ~\eqref{invariantansatz} as what follows.
 \begin{eqnarray}\label{explicitvalues}
 \begin{array}{cc}
 A_0=B_0=0, & B_1=\pm iB_2, \\
 A(t)=-\frac{iB_1}{\omega}e^{i\omega t}, &
 B(t)=\frac{B_1}{\omega}e^{i\omega t},  \\ 
 C(t)=-\frac{\eta}{2\hbar}B(t), &  D(t)=\frac{\eta}{2\hbar}A(t),  \\
 \alpha(t)=-\left(1+ \frac{\theta \eta}{4 \hbar^2}\right)\frac{m^2 \hbar^2}{\eta^2}B_1 e^{i\omega t}. &  
 \end{array}
 \end{eqnarray}
 
 The solution of ~\eqref{nuequationexplicit} then turns out to be
 \begin{eqnarray}
 \nu_\lambda (t)&=&\ln\nu_1 + i\ln\Phi_0 + \hbar_\eta^\theta t + \frac{i}{2B_1}\lambda e^{-i\omega t}.\\
 \mbox{With}\;\; \hbar_\eta^\theta &=& \frac{m^2g\theta}{2\eta} + \frac{m^2g\theta^2}{8\hbar^2} -\frac{\eta}{2m\hbar}.
 \end{eqnarray}
 Where $\ln\nu_1$ is the integration constant.\\
 Now we can assume the existence of annihilation operators $\hat{J}_1$ and $\hat{J}_2$ as follows.
 \begin{eqnarray}\label{anihilationdefn}
 \hat{J}_1&=& A(t)\hat{p}_x + C(t)\hat{x}, \;\;
 \hat{J}_2= B(t)\hat{p}_y + D(t)\hat{y}, \\
 \mbox{i.e}, \;\; \hat{\mathcal{I}}&=& \hat{J}_1+ \hat{J}_2 + \alpha(t)\hat{\mathbb{I}}.\nonumber
 \end{eqnarray}
 With the help of ~\eqref{explicitvalues}, One may readily verify that
 \begin{eqnarray}\label{annihilationcommutation}
 \left[\hat{J}_i, \hat{J}^\dag_j\right]&=& \frac{m^2\hbar^2}{\eta} \vert B_1 \vert ^2 \delta_{ij},\; i,j=1,2.\nonumber\\
 \left[\hat{J}_i, \hat{J}_j\right]&=&\left[\hat{J}^\dag_i, \hat{J}^\dag_j\right]=0. 
 \end{eqnarray}
 One can always choose the  parameter $B_1$ suitably. In particular, the choice 
 \begin{equation}\label{modB1}
 \vert B_1\vert=\frac{\sqrt{\eta}}{m\hbar},
 \end{equation}
 simplifies the algebra in ~\eqref{annihilationcommutation} as follows
 \begin{equation}
  \left[\hat{J}_i, \hat{J}^\dag_j\right]= \delta_{ij}.
  \end{equation}
 We have used the choice ~\eqref{modB1} in subsequent calculations. 
 To incorporate the all possible values of the free parameter $\lambda$ (the eigenvalues of $\hat{\mathcal{I}}$), one has to take a weighted sum of $\psi_\lambda$ as follows. 
 \begin{equation}\label{psidefn}
 \Psi (x,y,t)=\int\limits_{-\infty}^{+\infty}\psi_\lambda (x,y,t)\mu(\lambda) d\lambda .
 \end{equation}
 To ensure the convergence of the integration ~\eqref{psidefn} with ~\eqref{Phiform} and ~\eqref{psilambda}, it is sufficient to take a Gaussian type weight function $\mu(\lambda)$. In particular
 \begin{equation}\label{weightfunction}
 \mu(\lambda)=e^{-\frac{\kappa\lambda^2}{2\hbar}}.
 \end{equation}
 $\kappa \ge 0$ is a real parameter which is introduced in the weight function $\mu(\lambda)$ to ensure convergence of the integration ~\eqref{psidefn}. With the help of ~\eqref{weightfunction} in ~\eqref{psidefn} and using ~\eqref{Phiform} and ~\eqref{psilambda} we can obtain the explicit form of $\Psi(x,y,t)$, which reads
 \begin{eqnarray}
 \Psi (x,y,t)= \nu_0 \sqrt{\frac{2\pi\hbar}{\kappa}} \exp[ -\frac{\eta}{4\hbar^2}\vert z \vert ^2 -\frac{1}{2\hbar\omega}\left(1+\frac{\theta\eta}{4\hbar^2}\right)\bar{z}+ \nonumber \\ \frac{1}{8\hbar B_1^2 \kappa}\left(i\hbar-\omega\bar{z}\right)^2 e^{-2i\omega t}+i \hbar_\eta^\theta t ].
 \end{eqnarray}
 Here the notation $z=x+iy$ has been used.  $\bar{z}$ is the complex conjugate of $z$. $\nu_0$ is a constant. The probability density is given by 
 \begin{eqnarray}\label{rhodefn}
 \rho(x,y,t)=\vert \Psi(x,y,t)\vert ^2=\bar{\Psi }\Psi .
 \end{eqnarray}
 The normalizability  of $\psi$, i.e., the boundedness of ~\eqref{rhodefn} under usual $L^2(-\infty,+\infty)$ norm leads to the following restriction on the allowed values of $\kappa$.
 \begin{equation}
 \kappa \ge \frac{\eta}{2m^2\hbar b_1^2}=\frac{\hbar}{2}.
 \end{equation} 
 For simplicity and as well as for the later convenience, in subsequent computations, we shall use
 \begin{equation}
 \kappa = \frac{\hbar}{2}.
 \end{equation}

 \section{Expectation values of some observables}
 Expectation value of an observable $\hat{\mathcal{O}}$ on the state $\psi$ is defined by
 \begin{equation}
 \langle \hat{\mathcal{O}} \rangle _\psi = \left(\psi,\hat{\mathcal{O}}\psi\right).
 \end{equation}
 Here, $(,)$ denotes the scalar product. In our case we shall use the usual convenient scalar product via the integration.
 \begin{eqnarray}
 \langle \hat{x} \rangle &=& \frac{\vert \tilde{\nu}_0\vert^2 \pi}{4a_0 \sqrt{a_0\beta_0}}\left( h_0 \gamma_0 - c_0\right)e^{-2\delta_0}. \label{expectationx}\\
 \langle \hat{y} \rangle& =& \frac{\vert \tilde{\nu}_0\vert^2 \pi}{4b_0 \sqrt{b_0\beta_1}}\left( h_0 \gamma_1 - d_0\right)e^{-2\delta_1}. \\
 \langle \hat{p}_x \rangle& =& -\frac{\hbar}{i}\left(c+ 2a \langle \hat{x} \rangle +h \langle \hat{y} \rangle \right) . \label{expectationpx}\\
 \langle \hat{p}_y \rangle &=& -\frac{\hbar}{i}\left(d+ 2b \langle \hat{y} \rangle +h \langle \hat{x} \rangle \right). \\
 \langle \hat{x}^2 \rangle &=&\frac{\vert \tilde{\nu}_0\vert^2 \pi}{8 a_0 \sqrt{a_0\beta_0}} \left[1+ \frac{h_0^2}{4a_0\beta_0}+ \frac{1}{a_0}\left(c_0-h_0\gamma_0\right)^2\right]e^{-2\delta_0}. \label{expectationx2}\\
 \langle \hat{y}^2 \rangle &=&\frac{\vert \tilde{\nu}_0\vert^2 \pi}{8 b_0 \sqrt{b_0\beta_1}} \left[1+ \frac{h_0^2}{4b_0\beta_1}+ \frac{1}{b_0}\left(d_0-h_0\gamma_1\right)^2\right]e^{-2\delta_1} .\\
 \langle \hat{p}_x ^2\rangle &=&\hbar^2 \left[\left(2a-c^2\right)- 4ac \langle \hat{x} \rangle -2ch \langle \hat{y} \rangle -4a^2\langle \hat{x}^2 \rangle -h^2\langle \hat{y} ^2\rangle -4ah\langle \hat{x}\hat{y} \rangle \right] . \label{expectationpx2}\\
 \langle \hat{p}_y ^2\rangle &=&\hbar^2 \left[\left(2b-d^2\right)- 4bd \langle \hat{y} \rangle -2dh \langle \hat{x} \rangle -4b^2\langle \hat{y}^2 \rangle -h^2\langle \hat{x}^2\rangle -4bh\langle \hat{y}\hat{x} \rangle \right].\\
 \langle \hat{x} \hat{y}\rangle &=& \frac{\vert \tilde{\nu}_0\vert^2 \pi}{4a_0 \sqrt{a_0\beta_0}}\left( c_0 \gamma_0 - h_0\gamma_0^2 -\frac{h_0}{4\beta_0}\right)e^{-2\delta_0} .
 \end{eqnarray}
 Where $a_0,b_0,c_0,d_0, h_0 \;\mbox{and}\;k_0$ are Real part of $a,b,c,h,\;\mbox{and}\;k$ respectively. The values of $a,b,c,h,\;\mbox{and}\;k$ are given by the following.
 \begin{eqnarray}
 a&=&\frac{\eta}{4\hbar^2}\left(1-e^{-2i\omega\tau }\right),\;\;
 b=\frac{\eta}{4\hbar^2}\left(1+e^{-2i\omega\tau }\right),\nonumber \\
 c&=&\frac{m}{2\eta}\left(1+\frac{\theta\eta}{4\hbar^2}\right) + \frac{im}{2}e^{-2i\omega(t+\tau)},\;\;
 d=-\frac{im}{2\eta}\left(1+\frac{\theta\eta}{4\hbar^2}\right) + \frac{m}{2}e^{-2i\omega(t+\tau)},\nonumber \\
 h&=& \frac{i\eta}{2\hbar^2}e^{-2i\omega(t+\tau) },\;\;
 k= \frac{m^2\hbar^2}{4\eta}e^{-2i\omega(t+\tau) } -i\hbar_\eta^\theta t ,\;\;
 \tilde{\nu}_0=2\sqrt{\pi}\nu_0.\nonumber
 \end{eqnarray}
 The normalization factor $\vert \nu_0\vert$ is given by
 \begin{equation}
 \vert \nu_0\vert =\frac{1}{\pi}\left(\frac{a_0\beta_0}{4}\right)^{\frac{1}{4}}e^{\delta_0}.
 \end{equation}
 And $\beta_0,\beta_1,\gamma_0,\gamma_1,\delta_0,\delta_1$ are given by 
 \begin{eqnarray}
 \beta_0 &=&b_0-\frac{h_0^2}{4a_0}, \;\;
 \beta_1 =a_0-\frac{h_0^2}{4b_0},\nonumber \\
 \gamma_0 &=& \frac{d_0-c_0h_0}{2\left(b_0-\frac{h_0^2}{4a_0}\right)},\;\;
 \gamma_1 = \frac{c_0-d_0h_0}{2\left(a_0-\frac{h_0^2}{4b_0}\right)},\nonumber\\
 \delta_0 &=& \left( k_0-\frac{c_0^2}{4a_0}\right) - \frac{(d_0-c_0h_0)^2}{4\left(b_0-\frac{h_0^2}{4a_0}\right)}, \nonumber\\
 \delta_1 &=& \left( k_0-\frac{d_0^2}{4b_0}\right) - \frac{(c_0-d_0h_0)^2}{4\left(a_0-\frac{h_0^2}{4b_0}\right)}.
 \end{eqnarray}
 Using these relations one can have the expectation values in explicit form.\\
 In order to verify that the obtained state is a squeezed coherent state, we need to verify the uncertainty relation \cite{dong2013}.  Using ~\eqref{expectationx} and ~\eqref{expectationx2} one can see that 
 \begin{equation}\label{disperesex}
 \langle \hat{x}^2\rangle -\langle \hat{x}\rangle^2 = \frac{1}{4a_0}\left(1+\frac{h_0^2}{4a_0\beta_0}\right).
 \end{equation}
 Since  $\hat{x}$ and $\hat{y}$ are uncorrelated, we have
 \begin{equation}\label{correletationxy}
 \langle \hat{x}\rangle \langle \hat{y}\rangle = \langle \hat{x} \hat{y}\rangle .
 \end{equation}
 Using ~\eqref{expectationpx} and ~\eqref{expectationpx2} and ~\eqref{correletationxy}, one can write
 \begin{equation}\label{dispersepx}
 \langle \hat{p}_x^2\rangle -\langle \hat{p}_x\rangle^2 = 2\hbar^2 a. 
 \end{equation}
 Now, ~\eqref{disperesex} and ~\eqref{dispersepx} guide us to write
 \begin{equation}\label{deltaxdeltapx}
 \vert \Delta\hat{x}  \Delta\hat{p}_x \vert ^2 = \frac{\hbar^2}{4}f_\tau(t),
 \end{equation}
 where
 \begin{equation}\label{ftaut}
 f_\tau(t) = \frac{2\csc\omega\tau}{1-\csc^22\omega\tau \sin^22\omega(t+\tau)} .
 \end{equation}
 $f_\tau(t)$ has local minima at the points $t=\frac{n\pi}{4\omega}-\tau$, where $n$ is  an even number. In particular, \begin{equation}\label{ftaumin}
 \min f_\tau(t)= 2\csc\omega\tau\csc^22\omega\tau > 2\;\;\forall \tau.
 \end{equation}
 Therefore, we can conclude that
 \begin{equation}\label{uncert}
  \Delta\hat{x}  \Delta\hat{p}_x  > \frac{\hbar}{2},
 \end{equation}
 which confirms that the concerned state is indeed a squeezed coherent state.
 \section{conclusion}
 We have seen that a Lewis-Riesenfeld invariant operator exists for the gravitational well for non-commutative geometry. We have also discussed the uncertainty relations (UR) via the expectation values (EV) on the state.  In particular, we have shown that the measure of uncertainty in terms of the standard deviation of observables is time-dependent.  The time-dependent function in the UR, as well as  EV, has a minimum, which indicates that the states are squeezed coherent states in general.  Thus, one can expect that a time-dependent fluctuation in the fringe width will appear in an actual experiment with interferometry. However, in the classical limit (i.e, $(\theta\to 0, \; \eta\to 0)$), there will be a rapidly varying periodic function in phase.  The time dependency will be averaged out by the rapidly varying periodic function in phase and thus will have a stable fringe. Therefore this type of system may be a potential candidate to determine the values of the noncommutative parameters $\theta,\; \eta$. This can be done with the route stated in \cite{Gouba}, in which with the aid of position-dependent noncommutativity, the upper bounds of the parameter of noncommutativity and the parameter of deformation were evaluated in comparison with the experimental results.  Using the method prescribed in \cite{dongbook}, one can even obtain the constraints between the undetermined parameters using Bethe ansatz. This may be an interesting topic for further study in this regime.  
 \section{Acknowledgement}
 We are grateful to the anonymous referees for their valuable suggestions, which were very much helpful.

 \end{document}